\documentclass[12pt]{article}

\usepackage{bbm}
\usepackage{amssymb,amsmath}

\allowdisplaybreaks

\newcommand{\be}{\begin{equation}}
\newcommand{\ee}{\end{equation}}
\newcommand{\ba}{\begin{eqnarray}}
\newcommand{\ea}{\end{eqnarray}}
\newcommand{\no}{\nonumber\\}
\newcommand{\zz}{\mathbbm{Z}_2}
\newcommand{\zzz}{\mathbbm{Z}_3}
\newcommand{\zzzz}{\mathbbm{Z}_4}
\newcommand{\g}{\mathcal{G}}

\begin{document}

\title{
\normalsize \hfill UWThPh-2009-03 \\
\normalsize \hfill CFTP/09-025 \\*[8mm]
\LARGE Is $S_4$ the horizontal symmetry \\
of tri-bimaximal lepton mixing?
}

\author{
W.~Grimus,$^{(1)}$\thanks{E-mail: walter.grimus@univie.ac.at}
\
L.~Lavoura$^{(2)}$\thanks{E-mail: balio@cftp.ist.utl.pt}
\
and P.O.~Ludl$ \, ^{(1)}$\thanks{E-mail: a0301971@unet.univie.ac.at}
\\*[3mm]
$^{(1)} \! $
\small University of Vienna, Faculty of Physics \\
\small Boltzmanngasse 5, A--1090 Vienna, Austria
\\*[2mm]
$^{(2)} \! $
\small Technical University of Lisbon, Inst.\ Sup.\ T\'ecnico, CFTP \\
\small 1049-001 Lisbon, Portugal
}

\date{15 September 2009}

\maketitle

\begin{abstract}
We determine the symmetry groups under which the charged-lepton
and the Majorana-neutrino mass terms are invariant.
We note that those two groups always exist trivially,
\textit{i.e.}~independently of the presence
of any symmetries in the Lagrangian,
and that they always have the same form.
Using this insight,
we re-evaluate the recent claim
that,
whenever lepton mixing is tri-bimaximal,
$S_4$ is the minimal unique horizontal-symmetry
group of the Lagrangian of the lepton sector,
with $S_4$ being determined by the symmetries
of the lepton mass matrices.
We discuss two models for tri-bimaximal mixing which
serve as counterexamples to this claim.
With these two models
and some group-theoretical arguments we illustrate that
there is no compelling reason
for the uniqueness of $S_4$. 
\end{abstract}

\newpage

\section{Introduction}

The fermion mass and mixing problem is still unsolved
after decades of research.
The study of this problem has acquired a new impetus
after the experimental confirmation of neutrino oscillations,
through the resulting information on neutrino masses
and on lepton mixing---see~\cite{results} for recent fits to the data.
Remarkably,
lepton mixing seems to be well described by the lepton mixing matrix
\be
U_\mathrm{HPS} = \left( \begin{array}{ccc}
2 \left/ \sqrt{6} \right. & 1 \left/ \sqrt{3} \right. & 0 \\ 
- 1 \left/ \sqrt{6} \right. & 1 \left/ \sqrt{3} \right. &
- 1 \left/ \sqrt{2} \right. \\ 
- 1 \left/ \sqrt{6} \right. & 1 \left/ \sqrt{3} \right. &
1 \left/ \sqrt{2} \right.
\end{array} \right)
\times
\mathrm{diag} \left( 1, \, e^{i \beta_2 / 2}, \, e^{i \beta_3 / 2} \right),
\label{HPS}
\ee
where $\beta_2$ and $\beta_3$ are the so-called Majorana phases.
This conjecture of highly symmetric tri-bimaximal mixing (TBM)
was made by Harrison,
Perkins and Scott (HPS) in 2002~\cite{HPS}
and has strongly stimulated model building 
since then---for recent reviews see~\cite{models}.
It has also been hypothesized that the mixing problem might be decoupled
from the mass problem~\cite{prescient,hkv},
\textit{i.e.}~that a lepton mixing matrix $U = U_\mathrm{HPS}$
could result from a model with a horizontal-symmetry group $\g$ which,
however,
leaves the lepton masses arbitrary.

Recently,
it has been argued by C.S.~Lam~\cite{lam1,lam2,lam3} that
TBM uniquely determines $S_4$---the group of the permutations
of four objects and the symmetry group of the cube
and of the regular octahedron---as the minimal
family-symmetry group $\g$ of the leptons.\footnote{An
early precursor of a TBM model with horizontal-symmetry group $S_4$
can be found in~\cite{pakvasa};
more recent $S_4$ models for TBM are given in~\cite{s4papers-TBM}.
Other $S_4$ models
are found in~\cite{s4papers-nonTBM}.} 
Lam's argument was based on the idea that
the symmetries of the mass matrices $M_\ell$ and $M_\nu$
of the charged leptons and the light neutrinos,
respectively,
must reflect the symmetries of the underlying Lagrangian which leads,
without tuning of parameters,
to those mass matrices;
therefore,
the horizontal-symmetry group of the Lagrangian
must be the one generated by the symmetries of $M_\ell$
together with those of $M_\nu$.

In this paper,
we scrutinize this claim by first performing---in
section~\ref{symmetries of mass matrices}---a general study
of the symmetries of the lepton mass terms, 
\textit{i.e.}~allowing for a general mixing matrix $U$. 
We point out that those symmetries of the mass matrices
\emph{always exist,
always have the same group structure}
and must therefore be physically empty.
In section~\ref{assumptions} we proceed to outline
the arguments in~\cite{lam1,lam2,lam3} leading to $\g = S_4$.
We then present,
in section~\ref{natural TBM},
two models which predict TBM but do not conform to the idea
that the symmetries of $M_\ell$ and $M_\nu$
are also symmetries of the Lagrangian.
The conclusions of this paper are presented in section~\ref{concl}.

\section{The symmetries of the fermion mass matrices}
\label{symmetries of mass matrices}

We shall assume that the light left-handed neutrinos $\nu_L$
are Majorana particles.
The mass terms of the charged leptons $\ell$
and of the light neutrinos are then given by 
\be
\mathcal{L}_\mathrm{mass} = 
- \bar \ell_L M_\ell \ell_R
- \bar \ell_R M_\ell^\dagger \ell_L
+ {\textstyle \frac{1}{2}} \left(
\nu_L^T C^{-1} M_\nu \nu_L
- \overline{\nu_L} M_\nu^\ast C \overline{\nu_L}^T \right),
\ee
where $C$ is the charge-conjugation matrix
in Dirac space. The
matrix $M_\nu$ (in family space) is symmetric.
Since $M_\ell$ is not necessarily Hermitian,
it is convenient to work instead with
\be
H_\ell = M_\ell M_\ell^\dagger,
\ee
having in mind that the $\nu_L$ and $\ell_L$
are united in gauge-$SU(2)$ doublets.
We shall assume that there are only three charged fermions and light neutrinos.
The mass matrices are diagonalized as
\ba
U_\ell^\dagger H_\ell U_\ell
&=&
\mathrm{diag} \left( m_e^2, m_\mu^2, m_\tau^2 \right),
\label{ell}
\\
U_\nu^T M_\nu U_\nu
&=&
\mathrm{diag} \left( m_1, m_2, m_3 \right),
\label{nu}
\ea
where the $m_\alpha$ ($\alpha = e,\mu,\tau$)
and the $m_j$ ($j = 1, 2, 3$) are real,
non-negative and non-degenerate.
The lepton mixing matrix is given by
\be
U = U_\ell^\dagger U_\nu.
\ee

Denoting the columns of $U_\ell$ by $u_\alpha$,
\textit{i.e.}~$U_\ell = \left( u_e, u_\mu, u_\tau \right)$,
equation~(\ref{ell}) can be cast in the form
\be\label{eigen-ell}
H_\ell u_\alpha = m_\alpha^2 u_\alpha.
\ee
Since $U_\ell$ is unitary,
$u_\alpha^\dagger u_\beta = \delta_{\alpha\beta}$
and $\sum_\alpha u_\alpha u_\alpha^\dagger = \mathbbm{1}_3$,
where $\mathbbm{1}_n$ is the $n \times n$ unit matrix.
Now consider the $3 \times 3$ matrix
\be
S_\ell \left( \theta_e, \theta_\mu, \theta_\tau \right)
= \sum_\alpha e^{i \theta_\alpha} u_\alpha u_\alpha^\dagger.
\ee
Because of the unitarity of $U_\ell$,
\be
S_\ell \left( \theta_e, \theta_\mu, \theta_\tau \right)
S_\ell \left( \theta^\prime_e, \theta^\prime_\mu, \theta^\prime_\tau \right)
=
S_\ell \left( \theta_e + \theta^\prime_e, \theta_\mu + \theta^\prime_\mu,
\theta_\tau + \theta^\prime_\tau \right)
\ee
and
\be
S_\ell \left( \theta_e, \theta_\mu, \theta_\tau \right) u_\alpha = 
e^{i \theta_\alpha} u_\alpha.
\ee
Moreover,
\be
S_\ell^\dagger \! \left( \theta_e, \theta_\mu, \theta_\tau \right)
H_\ell \,
S_\ell \left( \theta_e, \theta_\mu, \theta_\tau \right)
= H_\ell.
\ee
This means that
\emph{$H_\ell$ has a $U(1) \times U(1) \times U(1)$ symmetry}.
Notice that \emph{this $U(1) \times U(1) \times U(1)$ symmetry of $H_\ell$
always exists---it is completely independent of the specific form of $H_\ell$}
and of the specific weak basis that we choose to work in.
The symmetry may be reduced to $U(1) \times U(1)$
if we make the additional requirement
$S_\ell \left( \theta_e, \theta_\mu, \theta_\tau \right) \in SU(3)$,
in which case we must restrict ourselves
to $\theta_e + \theta_\mu + \theta_\tau = 0$.

We proceed to study the symmetries of $M_\nu$.
We write $U_\nu = \left( u_1, u_2, u_3 \right)$.
Then $u_j^\dagger u_k = \delta_{jk}$
and $\sum_j u_j u_j^\dagger = \mathbbm{1}_3$.
The diagonalization equation~(\ref{nu}) means that
\be
M_\nu u_j = m_j u_j^\ast.
\ee
Defining
\be
S_\nu \left( a_1, a_2, a_3 \right) = \sum_j a_j u_j u_j^\dagger
\ee
with $a_{1,2,3} = \pm 1$,
it is obvious that
$S_\nu^2 \left( a_1, a_2, a_3 \right) = \mathbbm{1}_3$ and that
\be
S_\nu^T \! \left( a_1, a_2, a_3 \right)
M_\nu \,
S_\nu \left( a_1, a_2, a_3 \right) = M_\nu.
\ee
Therefore,
\emph{the Majorana mass matrix has $\zz \times \zz \times \zz$ symmetry.}
The existence of \emph{this symmetry
is completely independent of the specific form of $M_\nu$}.
We may require $S_\nu \left( a_1, a_2, a_3 \right) \in SU(3)$,
in which case we must impose the condition $a_1 a_2 a_3 = 1$
and the symmetry reduces to $\zz \times \zz$.

Let us dwell a bit longer
on the form of $S_\nu \left( a_1, a_2, a_3 \right)$ in the $SU(3)$ case.
If one chooses one of the $a_j$ to be positive
while the other two are negative,
one obtains the three matrices~\cite{lam1}
\be
\label{G}
G_j = - \mathbbm{1}_3 + 2 u_j u_j^\dagger.
\ee
These matrices have eigenvectors and eigenvalues given by 
\ba
G_j u_j &=& u_j,
\\
G_j u_k &=& - u_k \quad \mathrm{for} \ k \neq j.
\ea
The set $\left\{ \mathbbm{1}, G_1, G_2, G_3 \right\}$
of symmetries of $M_\nu$~\cite{lam1,lam2,lam3}
forms a Klein four-group~\cite{king-luhn},
with the properties
\ba
G_j^2 &=& \mathbbm{1}_3 \quad \forall j,
\\
G_j G_k = G_k G_j &=& G_l \quad
\mbox{for}\ j \neq k \neq l \neq j.
\ea
Klein's four-group
is Abelian and isomorphic to $\zz \times \zz$.

We stress that the symmetries of $H_\ell$ and $M_\nu$
discussed in this section are \emph{devoid of any physical content}.
They are \emph{mere mathematical consequences
of the diagonalizability of the mass matrices}.

\section{On the claim that $S_4$ is the minimal symmetry group of TBM}
\label{assumptions}

\subsection{The assumptions}
\label{lam}

It has been argued in~\cite{lam2,lam3} that
\begin{quote}
{\em The minimal finite family-symmetry group $\g$
yielding tri-bimaximal mixing (TBM) in the lepton sector
is uniquely given by $S_4$.}
\end{quote}
Let us
comment on the assumptions used in~\cite{lam2,lam3}.
Apart from taking for granted three lepton families
and assuming that the left-handed charged-lepton and neutrino fields
form doublets of the gauge group $SU(2)$,
a main ingredient is the Majorana nature of the 
light neutrinos.
This is actually an experimental question,
but there are many theoretical prejudices
in favour of the Majorana nature of neutrinos,
stemming from the seesaw
mechanism\footnote{On the other hand,
one must keep in mind the fact that
Dirac neutrino masses may also be suppressed
by one or more type~II seesaw mechanisms
acting on the vacuum expectation values
of Higgs doublets~\cite{branimir}.} and from Grand Unified Theories.
Of course, also TBM is an experimental
question;
a measurement of a non-zero $U_{e3}$ may be forthcoming in the near future
and would challenge TBM.\footnote{It is possible,
though,
that TBM holds at a high (seesaw) energy scale
but large deviations from it result,
if neutrino masses are almost degenerate,
from the renormalization-group evolution of mixing
down to the Fermi scale.}
On the group-theoretical side,
the main assumption~\cite{lam3} is that 
$\g$ is a finite subgroup of $U(3)$.
The left-handed-lepton gauge doublets are put in
a three-dimensional irreducible representation (irrep) of $\g$,
because it had been demonstrated in~\cite{lam1} 
that this is necessary to enforce TBM.
The finiteness of $\g$ is a simple means of avoiding Goldstone bosons,
but it excludes continuous horizontal-symmetry groups
like $SO(3)$~\cite{king} and $SU(3)$~\cite{ross}.

Since a horizontal- or family-symmetry group is responsible for TBM,
this means that TBM only holds at the tree level
and is therefore only approximate.
Indeed,
the different charged-lepton masses will always require a breaking of $\g$
and TBM will hence be modified by radiative corrections.
In~\cite{lam2,lam3}
the breaking of $\g$ was assumed to be spontaneous but,
as far as we can see,
soft breaking is also in accordance with the reasoning in those papers.
In that case,
$\g$ would be a symmetry only of the dimension-four terms
in the Lagrangian.

The most important ingredient of~\cite{lam1,lam2,lam3}
is the prescription for the determination of $\g$
from the symmetries of $M_\ell$ and $M_\nu$.
The details of that prescription
may be subsumed in the following way.
In the weak basis where $H_\ell$ is diagonal,
$U_\ell$ is diagonal as well and $U_\nu = U_\mathrm{HPS}$.\footnote{We
assume that a rephasing of the charged-lepton fields has been performed
in order to eliminate unphysical phases
which might otherwise be present to the left of $U_\mathrm{HPS}$.}
In that weak basis,
the matrices of equation~(\ref{G}) are given by  
\be
\label{GHPS}
G_1 = {\textstyle \frac{1}{3}} \left( \begin{array}{ccc}
1 & -2 & -2 \\ -2 & -2 & 1 \\ -2 & 1 & -2 \end{array} \right),
\quad
G_2 = {\textstyle \frac{1}{3}} \left( \begin{array}{ccc}
-1 & 2 & 2 \\ 2 & -1 & 2 \\ 2 & 2 & -1 \end{array} \right),
\quad
G_3 = \left( \begin{array}{ccc}
-1 & 0 & 0 \\ 0 & 0 & -1 \\ 0 & -1 & 0 \end{array} 
\right) 
\ee
and $S_\ell \left( \theta_e, \theta_\mu, \theta_\tau \right)$
is a diagonal phase matrix.
It is claimed in~\cite{lam1,lam2,lam3} that
\begin{quote}
{\em The horizontal-symmetry group $\g$ must be generated
by the three $G_j$ of equation~(\ref{GHPS})
together with only one matrix
$S_\ell \left( \theta_e, \theta_\mu, - \theta_e - \theta_\mu \right)
\in SU(3)$, 
with a specific choice of the phases $\theta_e$ and $\theta_\mu$
such that $\g$ turns out to be finite.}
\end{quote}
It had previously been stressed by other authors---see
for instance~\cite{hkv,feruglio1,blum}---that predictions for lepton mixing
hinge on the separate breaking of $\g$
to different subgroups in the charged-lepton sector
and in the neutrino sector.
The above claim amounts to saying that this also works in the opposite way:
the symmetries of $M_\nu$ may be put together with
\emph{one of} the symmetries of $H_\ell$ to uncover the minimal~$\g$.

In order to better appreciate the vast consequences
of this crucial point in the argumentation of~\cite{lam1,lam2,lam3},
we list some of its implications:
\begin{itemize}
\item The minimal group $\g$
is independent of the mechanism for small neutrino masses.
If,
for instance,
that mechanism is the type~I seesaw mechanism~\cite{seesaw}, 
then the number of right-handed neutrino singlets,
and the representation of $\g$ in which one chooses to place them,
have no bearing on the determination of $\g$ itself.
\item The same holds for the scalar content
of any theory with TBM at the tree level.
According to the above claim,
the number and transformation properties of the scalar fields under $\g$
are irrelevant for the determination of the minimal $\g$.
\item Implicitly,
the claim also implies that
any alignment of vacuum expectation values (VEVs) required for obtaining TBM
can be achieved without parameter tuning
in the model's scalar potential~\cite{lam2}.
\end{itemize}

\subsection{TBM and $S_4$} \label{s4tbm}

We next outline how one arrives at $S_4$ following the reasoning
in~\cite{lam1,lam2,lam3}. 
It is convenient to simplify the matrices $G_j$ of equation~(\ref{GHPS})
by transforming them to a different weak basis.
We define
\be
U_\omega = \frac{1}{\sqrt{3}} \left( \begin{array}{ccc}
1 & 1 & 1 \\
1 & \omega & \omega^2 \\
1 & \omega^2 & \omega
\end{array} \right)
\quad \mbox{with} \quad \omega = e^{2i\pi/3}
\ee
and perform the basis transformation 
$G_j \to \tilde G_j = U_\omega G_j U_\omega^\dagger$.
We find
\be
\tilde G_1 = \left( \begin{array}{ccc}
-1 & 0 & 0 \\ 0 & 0 & 1 \\ 0 & 1 & 0 \end{array} \right),
\quad
\tilde G_2 = \left( \begin{array}{ccc}
1 & 0 & 0 \\ 0 & -1 & 0 \\ 0 & 0 & -1 \end{array} \right),
\quad
\tilde G_3 = \left( \begin{array}{ccc}
-1 & 0 & 0 \\ 0 & 0 & -1 \\ 0 & -1 & 0 \end{array} \right).
\label{tgj}
\ee
In~\cite{lam2,lam3} it was argued that
the \emph{minimal} (finite) $\g$ leading to TBM
is generated by the matrices $G_j$ of equation~(\ref{GHPS})
together with
\be
\label{F}
F = S_\ell \left( 0, - \frac{2 \pi}{3}, \frac{2 \pi}{3} \right)
= \mathrm{diag} \left( 1, \omega^2, \omega \right).
\ee
We may transform $F$ as
\be
F \to \tilde F = U_\omega F U_\omega^\dagger
= \left( \begin{array}{ccc}
0 & 0 & 1 \\ 1 & 0 & 0 \\ 0 & 1 & 0
\end{array} \right).
\label{tf}
\ee
The group generated by $G_{1,2,3}$ and $F$
is the same as the one generated by $\tilde G_{1,2,3}$ and $\tilde F$.
It is clear from Appendix~A that
$\tilde F$ and the three $\tilde G_j$
generate the three-dimensional irrep $\mathbf{3}^\prime$ of $S_4$.
Its concrete realization is
\be
\label{assignment}
(12) \to \tilde G_1, \quad
(12)(34) \to \tilde G_2, \quad
(34) \to \tilde G_3, \quad
(234) \to \tilde F. 
\ee

\subsection{The groups generated by $\tilde F$ and one of the $\tilde G_j$}
\label{groupsG}

Let us for the time being assume that
the argumentation in~\cite{lam1,lam2,lam3} is sound
and that it indeed leads
to the minimal symmetry groups describing specific mixing cases.

We firstly consider bimaximal mixing,
\textit{i.e.}~the situation in which the third column of $U$ is
\be
u_3 = \frac{1}{\sqrt{2}} \left( \begin{array}{c}
0 \\ -1 \\ 1
\end{array} \right),
\ee
while the other two columns of $U$ are orthogonal to $u_3$
but otherwise arbitrary.
The minimal group leading to this situation
should be the one generated by $\tilde G_3$ and $\tilde F$;
this is,
following~(\ref{assignment}),
$S_3$---the permutation group of the numbers 2,
3 and 4.
Thus,
$S_3$ should be the minimal finite family-symmetry group of a model
which enforces bimaximal mixing without having recourse to tunings.

Let us secondly perform
the same exercise for trimaximal mixing,
\textit{i.e.}~for the situation in which the second column of $U$ is
\be
u_2 = \frac{1}{\sqrt{3}}
\left( \begin{array}{c} 1 \\ 1 \\ 1 \end{array} \right)
\ee
and the other two columns are arbitrary.
The symmetry group of this situation should be 
the one generated by $\tilde G_2$ and $\tilde F$.
According to~(\ref{assignment}) this is $A_4$---the group
of the even permutations of four numbers.

Finally let us investigate what is
the group generated solely by $\tilde G_1$ and $\tilde F$.
It is easy to convince oneself that this is the full $S_4$.
We reach the surprising conclusion that,
according to the reasoning in~\cite{lam1,lam2,lam3},
the minimal finite $\g$ of a model predicting
\be
\label{u1}
u_1 = \frac{1}{\sqrt{6}} \left( \begin{array}{c}
2 \\ -1 \\ -1 \end{array} \right)
\ee
is exactly the same $\g$ of a model predicting full TBM.

The findings of~\cite{lam2,lam3} outlined in subsection~\ref{s4tbm},
and their corollaries in this subsection,
on the minimal family-symmetry groups
capable of leading to specific mixing patterns,
come unexpected because they do not seem to be realized in existing models,
as can be noticed by visiting the literature
on bimaximal mixing---see~\cite{bi} for recent models
and the references therein for older ones---trimaximal
mixing~\cite{tri} and especially
tri-bimaximal mixing---see~\cite{hkv,feruglio1,feruglio,tbm}
and the reviews in~\cite{models}.
It also seems strange that models based on $S_4$ may,
depending on unspecified circumstances,
produce either TBM or simply equation~(\ref{u1});
notice that in the latter case the phenomenology of lepton mixing
can differ substantially from that of TBM~\cite{albright}.

\section{Two TBM models which contradict Lam's claim}
\label{natural TBM}

In this section we reconsider 
two renormalizable non-supersymmetric models,
one of them based on a horizontal-symmetry group $A_4$
and the other one on $S_4$,
which predict TBM without recourse to the tuning of parameters.
We focus on the symmetry structure of those models and,
in particular,
show that some of the symmetries of the mass matrices
are \emph{not} symmetries of the Lagrangian.
We do \emph{not} consider the scalar potentials that may,
in each case,
lead to the required vacuum states;
those issues have been addressed in the original papers.

\subsection{A model based on $A_4$}
\label{A4model}

The model of He,
Keum and Volkas (HKV)~\cite{hkv} has
horizontal symmetry $A_4$.
The finite group $A_4$
is generated by two transformations $A$ and $B$ satisfying 
(see \textit{e.g.}~\cite{genA4}) 
\be
A^2 = B^3 = \left( A B \right)^3 = e,
\label{a4}
\ee
where $e$ is the identity transformation.
It has three singlet irreps
\be
\mathbf{1}_j: \quad A \to 1, \ B \to \omega^j,
\ee
for $j = 0, 1, 2$
($\mathbf{1}_0$ is the trivial representation).
The only faithful irrep of $A_4$ is
\be
\mathbf{3}: \quad
A \to G_2,
\
B \to F.
\label{3}
\ee
As we will see,
this basis of the $\mathbf{3}$
has the advantage of leading to a diagonal $M_\ell$ in the HKV model; 
note that in the original paper~\cite{hkv} a different basis was used.
If $\left( a_1, b_1, c_1 \right)$ and $\left( a_2, b_2, c_2 \right)$
each transform as a $\mathbf{3}$ of $A_4$,
then~\cite{genA4}
\ba
a_1 a_2 + b_1 c_2 + b_2 c_1 \ \ \mathrm{is\ a\ } \mathbf{1}_0, & & \\
b_1 b_2 + a_1 c_2 + a_2 c_1 \ \ \mathrm{is\ a\ } \mathbf{1}_1, & & \\
c_1 c_2 + a_1 b_2 + a_2 b_1 \ \ \mathrm{is\ a\ } \mathbf{1}_2, & & \\
\left( 2 a_1 a_2 - b_1 c_2 - b_2 c_1, \
2 c_1 c_2 - a_1 b_2 - a_2 b_1, \
2 b_1 b_2 - a_1 c_2 - a_2 c_1 \right)
\ \ \mathrm{is\ a\ } \mathbf{3}. & &
\ea

The HKV model is a type-I-seesaw model
with three right-handed neutrinos $\nu_{jR}$ ($j = 1, 2, 3$).
The scalar sector comprehends
four Higgs doublets $\phi_k$ $(k = 0, 1, 2, 3$)
and three complex gauge singlets $\chi_j$.
The HKV model has an auxiliary symmetry $\zz$ under which
the $\nu_{jR}$ and $\phi_0$ change sign;\footnote{In
the original paper~\cite{hkv}
HKV actually used a $U(1)$ auxiliary symmetry.
The exact form of the auxiliary symmetry is,
however,
immaterial for our purposes here.}
the purpose of this $\zz$ is
to allow the right-handed neutrinos to have Yukawa couplings
only to the doublet $\phi_0$.
The multiplets of $A_4 \times \zz$ used in the HKV model
are given in table~\ref{hkvtable}
(the $D_{\alpha L}$ are the left-handed-lepton gauge doublets).
\begin{table}
\begin{center}
\begin{tabular}{|c|c|c|}
\hline
irrep & $A_4$ & $\zz$ \\
\hline \hline
$\left( \bar D_{eL}, \bar D_{\tau L}, \bar D_{\mu L} \right)$ &
$\mathbf{3}$ & $1$ \\
$\left( \nu_{1R}, \nu_{2R}, \nu_{3R} \right)$ & $\mathbf{3}$ & $-1$ \\
$e_R$ & $\mathbf{1}_0$ & $1$ \\
$\mu_R$ & $\mathbf{1}_2$ & $1$ \\
$\tau_R$ & $\mathbf{1}_1$ & $1$ \\
\hline \hline
$\phi_0$ & $\mathbf{1}_0$ & $-1$ \\
$\left( \phi_1, \phi_2, \phi_3 \right)$ & $\mathbf{3}$ & $1$ \\
$\left( \chi_1, \chi_2, \chi_3 \right)$ & $\mathbf{3}$ & $1$ \\
\hline \hline
\end{tabular}
\end{center}
\caption{Multiplets of the HKV model.
\label{hkvtable}}
\end{table}
The Majorana mass term of the right-handed neutrino singlets is 
\be
\mathcal{L}_\mathrm{Majorana} =
- \frac{m}{2} \left( \bar \nu_{1R} C \bar \nu_{1R}^T
+ \bar \nu_{2R} C \bar \nu_{3R}^T + \bar \nu_{3R} C \bar \nu_{2R}^T \right)
+ \mathrm{H.c.}
\ee
The Lagrangian of Yukawa couplings is
\ba
\mathcal{L}_\mathrm{Yukawa} &=&
- y_1 \left(
\bar D_{eL} \nu_{1R} + \bar D_{\mu L} \nu_{2R} + \bar D_{\tau L} \nu_{3R}
\right) \left( i \tau_2 \phi_0^\ast \right)
\no & &
- y_2 \left(
\bar D_{eL} \phi_1 + \bar D_{\mu L} \phi_2 + \bar D_{\tau L} \phi_3
\right) e_R
\no & &
- y_3 \left(
\bar D_{\tau L} \phi_2 + \bar D_{eL} \phi_3 + \bar D_{\mu L} \phi_1
\right) \mu_R
\no & &
- y_4 \left(
\bar D_{\mu L} \phi_3 + \bar D_{eL} \phi_2 + \bar D_{\tau L} \phi_1
\right) \tau_R
\no & &
- {\textstyle \frac{1}{2}}
\left( y_5 \chi_1 + y_6 \chi_1^\ast \right)
\left( 2 \bar \nu_{1R} C \bar \nu_{1R}^T
- \bar \nu_{2R} C \bar \nu_{3R}^T - \bar \nu_{3R} C \bar \nu_{2R}^T \right)
\no & &
- {\textstyle \frac{1}{2}}
\left( y_5 \chi_3 + y_6 \chi_2^\ast \right)
\left( 2 \bar \nu_{2R} C \bar \nu_{2R}^T
- \bar \nu_{1R} C \bar \nu_{3R}^T - \bar \nu_{3R} C \bar \nu_{1R}^T \right)
\no & &
- {\textstyle \frac{1}{2}}
\left( y_5 \chi_2 + y_6 \chi_3^\ast \right)
\left( 2 \bar \nu_{3R} C \bar \nu_{3R}^T
- \bar \nu_{1R} C \bar \nu_{2R}^T - \bar \nu_{2R} C \bar \nu_{1R}^T \right)
+ \mathrm{H.c.}
\label{LA4}
\ea
Let $v_k$ denote the VEV of $\phi_k^0$.
The neutrino Dirac mass matrix is proportional to the unit matrix:
$M_D = y_1^\ast v_0 \mathbbm{1}_3$.
If \emph{$v_2$ and $v_3$ vanish},
then the charged-lepton mass matrix is diagonal,
with $m_e = \left| y_2 v_1 \right|$,
$m_\mu = \left| y_3 v_1 \right|$ and $m_\tau = \left| y_4 v_1 \right|$.
On the other hand,
if \emph{the VEVs of the $\chi_j$ are all equal},
\textit{i.e.}~if 
$\left\langle \chi_1 \right\rangle_0 = \left\langle \chi_2 \right\rangle_0
= \left\langle \chi_3 \right\rangle_ 0 \equiv u$,
then the right-handed-neutrino Majorana mass matrix is
\be
M_R = \left( \begin{array}{ccc}
m + 2 m^\prime & - m^\prime & - m^\prime \\
- m^\prime & 2 m^\prime & m - m^\prime \\
- m^\prime & m - m^\prime & 2 m^\prime
\end{array} \right),
\ee
where $m^\prime \equiv y_5 u + y_6 u^\ast$.
Using $M_D \propto \mathbbm{1}_3$
and the seesaw formula $M_\nu = - M_D^T M_R^{-1} M_D$,
we find that the effective light-neutrino
mass matrix\footnote{Incidentally,
the same mass matrix $M_\nu$ following from the HKV model
has also been obtained in two other and totally different
$A_4$ models~\cite{feruglio,babu}.} $M_\nu \propto M_R^{-1}$.
Since $M_R$ is of the required form to generate TBM,
$M_R^{-1}$ is of that form too,
so this model predicts TBM
as long as the vacuum state is of the assumed
form.\footnote{In~\cite{hkv} it has been shown that
there is a range of the parameters of the scalar potential
such that the desired VEVs constitute a global minimum,
provided $CP$ is conserved.}

We now ponder whether $S_4$ might be a symmetry group of the HKV model.
Since the charged-lepton mass matrix is diagonal,
we may directly apply the reasoning of subsections~\ref{lam} and~\ref{s4tbm}.
The matrix $M_R$ is invariant under
$M_R \to G^T\! M_R G$
for either $G = G_2$ or $G = - G_3$.
The matrix $H_\ell$ is invariant under $H_\ell \to F^\dagger H_\ell F$.
According to equation~(\ref{3}) and table~\ref{hkvtable},
$G_2$ and $F$ may be extended to the transformations $A$ and $B$,
respectively,
which generate the symmetry group $A_4$ of the Lagrangian.
Since $G_2$ together with $- G_3$ and $F$
generate the irrep $\mathbf{3}$ of $S_4$,
we only have to check whether one may extend $- G_3$
to a symmetry of the whole Lagrangian.
The obvious extension of $- G_3$ is
\be
G: \quad \bar D_{\mu L} \leftrightarrow \bar D_{\tau L}, \
\nu_{2R} \leftrightarrow \nu_{3R}, \
\mu_R \leftrightarrow \tau_R, \
\phi_2 \leftrightarrow \phi_3, \
\chi_2 \leftrightarrow \chi_3.
\ee
If this $G$ were a symmetry of the Lagrangian
then the HKV model would indeed possess full $S_4$ symmetry.
However,
we see from equation~(\ref{LA4}) that this would necessitate
$y_3 = y_4$ and,
as a consequence,
$m_\mu = m_\tau$,
which is clearly unacceptable.
Therefore $- G_3$ cannot be extended
to become a symmetry of the HKV Lagrangian
and the HKV model possesses $A_4$ but not $S_4$ symmetry.
The symmetries of the mass matrix $M_\nu$
are indeed symmetries of the terms with coefficients $m$,
$y_5$ and $y_6$ in the Lagrangian,
but they are \emph{not} and cannot be symmetries of the full Lagrangian.
This constitutes a counterexample to the argument in~\cite{lam2,lam3}.

\subsection{A model based on $S_4$}
\label{S4model}

This model was proposed in~\cite{tbm}.
In its original version the symmetry group used was quite large.
We shall present here a simplified version
based on a horizontal-symmetry group $S_4 \times \zzzz$.

This is a  type-I-seesaw model
with {\em five} right-handed neutrinos
$\nu_{1R}, \ldots, \nu_{5R}$;
these are placed in a $\mathbf{3}$ and a $\mathbf{2}$ of $S_4$.
There are four Higgs doublets $\phi_0, \ldots, \phi_3$
and \emph{one complex} scalar singlet $\chi$,
the two real components of which form a $\mathbf{2}$ of $S_4$.
We use for the irreps of $S_4$ the bases
given in Appendix~A.
The multiplets of the model are given in table~\ref{gltable}.
\begin{table}
\begin{center}
\begin{tabular}{|c|c|c|}
\hline
irrep & $S_4$ & $\zzzz$ \\
\hline \hline
$\left( \bar D_{eL}, \bar D_{\mu L}, \bar D_{\tau L} \right)$ &
$\mathbf{3}$ & $1$ \\
$\left( \nu_{1R}, \nu_{2R}, \nu_{3R} \right)$ & $\mathbf{3}$ & $1$ \\
$\left( \nu_{4R}, \nu_{5R} \right)$ & $\mathbf{2}$ & $i$ \\
$\left( e_R, \mu_R, \tau_R \right)$ & $\mathbf{3}$ & $-1$ \\
\hline \hline
$\phi_0$ & $\mathbf{1}$ & $1$ \\
$\phi_1$ & $\mathbf{1}$ & $-1$ \\
$\left( \phi_2, \phi_3 \right)$ & $\mathbf{2}$ & $-1$ \\
$\left( \chi, \chi^\ast \right)$ & $\mathbf{2}$ & $-1$ \\
\hline \hline
\end{tabular}
\end{center}
\caption{Multiplets of the $S_4$-based model.
\label{gltable}}
\end{table}

A crucial feature of the model
is the soft breaking of the horizontal symmetries.
The dimension-4 terms in the Lagrangian preserve $S_4$.
The dimension-3 terms break $S_4$ softly to its subgroup $S_3$.
The dimension-2 terms break $S_3$ softly to its subgroup $S_2$,
the $\mu$--$\tau$ interchange symmetry. 
This $S_2$ is broken only spontaneously {\em at the Fermi scale}.
Indeed,
at the seesaw scale $\chi$ gets a {\em real} VEV
$\left\langle \chi \right\rangle_0 \equiv u$,
which does not break the $S_2$ symmetry $\chi \leftrightarrow \chi^\ast$.
At the Fermi scale,
on the other hand,
$\phi_2$ and $\phi_3$ acquire different VEVs,
thereby breaking the $\mu$--$\tau$ interchange symmetry.

The auxiliary symmetry $\zzzz$ is softly broken
already by the terms of dimension three.

The Yukawa couplings have dimension four
and therefore respect the full horizontal symmetry:
\ba\label{yuk}
\mathcal{L}_\mathrm{Yukawa} &=&
- y_1 \left(
\bar D_{eL} \nu_{1R} + \bar D_{\mu L} \nu_{2R} + \bar D_{\tau L} \nu_{3R}
\right) \left( i \tau_2 \phi_0^\ast \right)
\no & &
- y_2 \left(
\bar D_{eL} e_R + \bar D_{\mu L} \mu_R + \bar D_{\tau L} \tau_R
\right) \phi_1
\no & &
- y_3 \left[
\left(
\bar D_{eL} e_R + \omega^2 \bar D_{\mu L} \mu_R + \omega \bar D_{\tau L} \tau_R
\right) \phi_2
\right. \no & & \left. +
\left(
\bar D_{eL} e_R + \omega \bar D_{\mu L} \mu_R + \omega^2 \bar D_{\tau L} \tau_R
\right) \phi_3
\right]
\no & &
- \frac{y_4}{2} \left( \bar \nu_{4R} C \bar \nu_{4R}^T \chi^\ast
+ \bar \nu_{5R} C \bar \nu_{5R}^T \chi \right)
+ \mathrm{H.c.}
\ea
The symmetry $\zzzz$ forbids Yukawa couplings
of the $\phi_j$ 
as well as of $\chi$ and $\chi^\ast$
to the $\nu_{kR}$ for $j,k = 1, 2, 3$.
The charged-lepton Yukawa couplings
are flavour-diagonal because there are no Higgs doublets in triplets of $S_4$.
When the Higgs doublets get VEVs
$\left\langle \phi_k^0 \right\rangle_ 0 = v_k$,
the charged leptons acquire masses:
\ba
m_e &=& \left| y_2 v_1 + y_3 \left( v_2 + v_3 \right) \right|,
\\
m_\mu &=& 
\left| y_2 v_1 + y_3 \left( \omega^2 v_2 + \omega v_3 \right)
\right|,
\\
m_\tau &=& 
\left| y_2 v_1 + y_3 \left( \omega v_2 + \omega^2 v_3 \right)
\right|.
\ea
Since the $\mu$--$\tau$ interchange symmetry 
is broken at the Fermi scale,
$v_2 \neq v_3$ and
this leads to $m_\mu$ and $m_\tau$ being different.

The neutrino Dirac mass matrix $M_D$
is a $5 \times 3$ matrix;
its upper $3 \times 3$ block is proportional to the unit matrix,
with proportionality coefficient
$y_1^\ast v_0$;
the lower $2 \times 3$ block is a null matrix.

The Majorana mass terms of the right-handed neutrinos
have dimension three and therefore respect $S_3$ but not $S_4$.
They are
\ba
\mathcal{L}_\mathrm{Majorana} &=&
- \frac{M_0}{2} \left( \bar \nu_{1R} C \bar \nu_{1R}^T
+ \bar \nu_{2R} C \bar \nu_{2R}^T
+ \bar \nu_{3R} C \bar \nu_{3R}^T \right)
\no & &
- M_1
\left( \bar \nu_{1R} C \bar \nu_{2R}^T
+ \bar \nu_{2R} C \bar \nu_{3R}^T
+ \bar \nu_{3R} C \bar \nu_{1R}^T \right)
\no & &
- M_2 \bar \nu_{4R} C \bar \nu_{5R}^T
\no & &
- M_3 \left[
\left( \bar \nu_{1R} + \omega^2 \bar \nu_{2R} + \omega \bar \nu_{3R} \right)
C \bar \nu_{5R}^T
\right. \no & & \left. +
\left( \bar \nu_{1R} + \omega \bar \nu_{2R} + \omega^2 \bar \nu_{3R} \right)
C \bar \nu_{4R}^T
\right]
+ \mathrm{H.c.}
\ea
(We remind the reader 
that the symmetry $\zzzz$ is broken softly at dimension three.)
The resulting $5 \times 5$ right-handed-neutrino Majorana mass matrix is
\be
M_R = \left( \begin{array}{ccccc}
M_0 & M_1 & M_1 & M_3 & M_3 \\
M_1 & M_0 & M_1 & \omega M _3 & \omega^2 M_3 \\
M_1 & M_1 & M_0 & \omega^2 M_3 & \omega M_3 \\
M_3 & \omega M_3 & \omega^2 M_3 & y_4 u & M_2 \\
M_3 & \omega^2 M_3 & \omega M_3 & M_2 & y_4 u
\end{array} \right).
\label{mr}
\ee
It is easy to convince oneself that $M_\nu = - M_D^T M_R^{-1} M_D$
has the structure to be diagonalized by $U_\mathrm{HPS}$~\cite{tbm}.

One can check that $\hat G_j^T  M_R \hat G_j = M_R$, 
with $5 \times 5$ real orthogonal matrices $\hat G_j$ given by 
\be
\hat G_1 = 
\left( \begin{array}{cc} G_1 & 0 \\ 0 & \tau_1 \end{array} \right),
\quad
\hat G_2 =
\left( \begin{array}{cc} - G_2 & 0 \\ 0 & \mathbbm{1}_2 \end{array} \right),
\quad
\hat G_3 = 
\left( \begin{array}{cc} - G_3 & 0 \\ 0 & \tau_1 \end{array} \right),
\label{G5}
\ee
where $0$ denotes the $3 \times 2$ or $2 \times 3$ null matrix
and $\tau_1$ is the first Pauli matrix.
The matrices $\hat G_j$ form
a Klein four-group.
It is easy to see that the last transformation in equation~(\ref{G5}),
\textit{viz.}
\be
\nu_{2R} \leftrightarrow \nu_{3R}, \quad
\nu_{4R} \leftrightarrow \nu_{5R}, \quad
D_{\mu L} \leftrightarrow D_{\tau L}, \quad
\mu_R \leftrightarrow \tau_R, \quad
\phi_2 \leftrightarrow \phi_3, \quad
\chi \leftrightarrow \chi^\ast
\ee
is a symmetry of the Lagrangian
(it is indeed part of its defining symmetry group $S_4$),
but the first two transformations in equation~(\ref{G5})
\emph{cannot} be extended to symmetries of the 
full Lagrangian---for a mathematical proof
see Appendix~B.
Therefore we find in this model that,
once again,
the symmetries of the mass matrices are not symmetries of the Lagrangian.

\section{Conclusions} \label{concl}

We summarize here the arguments that we have found
against $S_4$ being the unique horizontal-symmetry group
for tri-bimaximal mixing:
\begin{itemize}
\item Using exactly the same arguments as employed in~\cite{lam2,lam3}, 
one would find---as demonstrated in section~\ref{groupsG}---that
$S_4$ is the horizontal-symmetry group of any 
model leading to a lepton mixing matrix
whose first column $u_1$ is given by equation~(\ref{u1}), 
contradicting the claim that $S_4$ is the
symmetry of the \emph{full} TBM.
\item In section~\ref{A4model}
we have reconsidered the HKV model~\cite{hkv},
which has a horizontal-symmetry group $A_4$ 
generated by the $G_2$ in equation~(\ref{GHPS})
and the $F$ of equation~(\ref{F}).
We have found that in that model
the symmetry $G_3$ of the mass matrix $M_\nu$
is not a symmetry of the full Lagrangian.
Therefore,
the HKV model is a true $A_4$ model
and not an $S_4$ model,
in contradiction with the claim of~\cite{lam2,lam3}. 
\item In section~\ref{S4model} we have rewritten the model of~\cite{tbm}
in terms of a horizontal-symmetry group $S_4$.
However,
this symmetry group $S_4$ is not realized in the way
described in~\cite{lam2,lam3},
because the symmetry $\hat G_2$
[see equation~(\ref{G5})]
of $M_\nu$ is not a symmetry of the full Lagrangian.
\end{itemize}

There is precisely one reason why the argumentation in~\cite{lam2,lam3} fails: 
the symmetries of the mass matrices
are not in general residues
of the complete horizontal-symmetry group
of the Lagrangian. 
Specific properties of the mass matrices
are determined by the symmetries in the Lagrangian, 
but the symmetries of the mass matrices always exist trivially
and are nothing more than expressions of their diagonalizability.

We also want to emphasize that
it may happen that a TBM model can be interpreted
in terms of different horizontal-symmetry groups,
provided they all have the irreps and Clebsch--Gordan coefficients
needed for the construction of the model.
An example for this is the model of~\cite{tbm} in its original version;
although the different possible horizontal-symmetry
groups---the simplest of which is an extension of $\Delta(54)$---do not
all lead to exactly the same Lagrangian,
the terms in which they differ
reside exclusively in the scalar potential and are irrelevant for TBM.
Thus,
the group $S_4$ is not special for TBM,
it is simply one of many groups with which TBM models can be constructed.

\paragraph{Acknowledgements:}
W.G.\ and L.L.\ acknowledge support from the European Union
through the network programme MRTN-CT-2006-035505.
The work of L.L.~was supported by the Portuguese
\textit{Funda\c c\~ao para a Ci\^encia e a Tecnologia}
through the project U777--Plurianual.

\setcounter{equation}{0}
\renewcommand{\theequation}{A\arabic{equation}}

\section*{Appendix A} \label{s4}

Since $S_4$ figures prominently in~\cite{lam1,lam2,lam3},
we discuss its structure in detail.
Consider the two sets of matrices
\ba
K &=& \left\{
\mathbbm{1}_3, \
\mathrm{diag} \left( 1, -1, -1 \right), \
\mathrm{diag} \left( -1, 1, -1 \right), \
\mathrm{diag} \left( -1, -1, 1 \right)
\right\},
\\
S &=& \left\{
\mathbbm{1}_3, \
\left( \begin{array}{ccc}
0 & 1 & 0 \\ 0 & 0 & 1 \\ 1 & 0 & 0
\end{array} \right), \
\left( \begin{array}{ccc}
0 & 0 & 1 \\ 1 & 0 & 0 \\ 0 & 1 & 0
\end{array} \right), \
\right. \no & & \left.
\left( \begin{array}{ccc}
1 & 0 & 0 \\ 0 & 0 & 1 \\ 0 & 1 & 0
\end{array} \right), \
\left( \begin{array}{ccc}
0 & 0 & 1 \\ 0 & 1 & 0 \\ 1 & 0 & 0
\end{array} \right), \
\left( \begin{array}{ccc}
0 & 1 & 0 \\ 1 & 0 & 0 \\ 0 & 0 & 1
\end{array} \right)
\right\}.
\ea
Obviously,
$K$ is a representation of Klein's four-group
and $S$ is the defining reducible representation of $S_3$,
the permutation group of three elements.
We note that
\be
s k s^{-1} \in K \quad
\forall k \in K, s \in S.
\ee
$S_4$ may be viewed as the semidirect product $K \rtimes S$~\cite{bovier},
\textit{i.e.}
\be
S_4 = \left\{ \left. \left( k, s \right) \ \right| \
k \in K, s \in S \right\},
\ee
with the usual multiplication rule for semidirect products:
\be
\left( k_1, s_1 \right) \left( k_2, s_2 \right)
= \left( k_1 s_1 k_2 s_1^{-1}, s_1 s_2 \right).
\label{multip}
\ee
Since $K$ has four elements and $S$ has six elements,
$S_4$ has $4 \times 6 = 24$ elements.
The structure $S_4 \cong K \rtimes S_3$ implies 
$S_4 \cong \Delta \left( 24 \right)$~\cite{bovier}
and $S_3 \cong S_4 / K$.
Therefore,
all the irreps of $S_3$ can be extended to irreps of $S_4$.
In particular,
$S_4$ has the doublet irrep
\be
\mathbf{2}: \quad \left( k, s \right) \to D_2 \left( s \right),
\ee
where $D_2 \left( s \right)$ is the doublet irrep of $S_3$,
namely
\be
\left( \begin{array}{ccc}
1 & 0 & 0 \\ 0 & 0 & 1 \\ 0 & 1 & 0
\end{array} \right)
\to
\left( \begin{array}{cc} 0 & 1 \\ 1 & 0 \end{array} \right),
\quad
\left( \begin{array}{ccc}
0 & 0 & 1 \\ 0 & 1 & 0 \\ 1 & 0 & 0
\end{array} \right)
\to
\left( \begin{array}{cc} 0 & \omega^2 \\ \omega & 0 \end{array} \right),
\ee
where $\omega \equiv \exp{\left( 2 i \pi / 3 \right)}$.
The group $S_4$ also has,
besides the trivial representation $\mathbf{1}$,
another singlet irrep
\be
\mathbf{1}^\prime: \quad \left( k, s \right) \to \det{s},
\ee
which is also an irrep of $S_3$.
Finally,
$S_4$ has two triplet irreps,
\ba
\mathbf{3}: & & \left( k, s \right) \to k s,
\\
\mathbf{3}^\prime: & & \left( k, s \right) \to k s \left( \det{s} \right).
\ea
All the matrices of the $\mathbf{3}^\prime$
have determinant $+1$ and belong to $SO(3)$.
By contrast,
the matrices of the $\mathbf{3}$
which represent odd permutations of $S_4$
have determinant $-1$.

Since $1^2 + 1^2 + 2^2 + 3^2 + 3^2 = 24$,
these are all the irreps of $S_4$.

One may use the matrices of the $\mathbf{3}$
to represent the permutations of the four numbers 1,
2,
3 and 4 in the following way:
\be
\begin{array}{c}
\left( 1 2 \right) \left( 3 4 \right)
\to \mathrm{diag} \left( 1, -1, -1 \right),
\\
\left( 1 3 \right) \left( 2 4 \right)
\to \mathrm{diag} \left( -1, 1, -1 \right),
\\
\left( 1 4 \right) \left( 2 3 \right)
\to \mathrm{diag} \left( -1, -1, 1 \right),
\\
\left( 3 4 \right) \to
\left( \begin{array}{ccc}
1 & 0 & 0 \\ 0 & 0 & 1 \\ 0 & 1 & 0
\end{array} \right),
\quad
\left( 2 4 \right) \to
\left( \begin{array}{ccc}
0 & 0 & 1 \\ 0 & 1 & 0 \\ 1 & 0 & 0
\end{array} \right),
\quad
\left( 2 3 \right) \to 
\left( \begin{array}{ccc}
0 & 1 & 0 \\ 1 & 0 & 0 \\ 0 & 0 & 1
\end{array} \right).
\end{array}
\ee
It is clear in this representation that the matrices of $S$
represent the permutation group $S_3$ of the numbers 2,
3 and 4.
The matrices of $K$ allow one to additionally represent
the permutations of $S_4$ which involve the number 1.

As a side remark,
from the above discussion of $S_4$ it also follows that
$A_4 \cong K \rtimes \zzz \cong \Delta(12)$,
where $A_4$ is the group of the even permutations of four numbers.

\setcounter{equation}{0}
\renewcommand{\theequation}{B\arabic{equation}}
\section*{Appendix B}

Let $D_L$,
$\ell_R$ and $\phi$ denote the column vectors formed by,
respectively,
the fields $D_{\alpha L}$  ($\alpha = e, \mu, \tau$),
$\alpha_R$ and $\phi_j$ ($j = 1, 2, 3$).
We want to prove that the tranformation $D_L \to G_2 D_L$,
with $G_2$ given in equation~(\ref{GHPS}),
cannot be extended to $\ell_R$ and $\phi$ 
in such a way that it constitutes a symmetry of the Yukawa couplings
in equation~(\ref{yuk}).
Those couplings may be written 
\be
\label{yukl}
\bar D_L \left( \sum_j \Gamma_j \phi_j \right) \ell_R,
\ee
with diagonal coupling matrices
\be
\Gamma_1 = y_2 \mathbbm{1}_3, \quad 
\Gamma_2 = y_3\, \mbox{diag} \left( 1, \omega^2, \omega \right), \quad
\Gamma_3 = y_3\, \mbox{diag} \left( 1, \omega, \omega^2 \right).
\ee
We assume that
there is a symmetry transformation of the Yukawa couplings~(\ref{yukl})
given by
\be
\label{trafo}
D_L \to G_2 D_L, \quad \ell_R \to P \ell_R, \quad \phi \to Q \phi,
\ee
with $3 \times 3$ unitary matrices $P$ and $Q$.
Then we obtain the invariance condition
\be
\sum_j \left( G_2 \Gamma_j P \right) Q_{jk} = \Gamma_k. 
\ee
Shifting $Q$
to the right-hand side and multiplying the resulting
equation with its Hermitian conjugate,
we arrive at
\be
G_2 \Gamma_p \Gamma_q^\dagger G_2 = 
\sum_{j,k} \Gamma_j \Gamma_k^\dagger\, Q_{pj}^\ast Q_{qk}.
\ee
Since the right-hand side of this equation
is a sum of diagonal matrices,
we find that $G_2 \Gamma_p \Gamma_q^\dagger G_2$ is diagonal
for all indices $p$ and $q$.
It is easy to check that this is possible only if $y_3 = 0$,
which leads to degenerate charged leptons.
Therefore,
no reasonable symmetry~(\ref{trafo}) exists.

\end{document}